\documentclass[
 reprint,
 amsmath,amssymb,
 aps,
 floatfix,
 prapplied
]{revtex4-2}

\usepackage{subfigure}
\usepackage{graphicx}
\usepackage{dcolumn}
\usepackage{bm}
\usepackage{appendix}
\usepackage[dvipsnames]{xcolor}

\begin{document}


\title{Analysis of ion chain sympathetic cooling and gate dynamics}

\author{A. Paul}
 \altaffiliation[Currently at ]{Department of Electrical Engineering and Computer Science, Massachusetts Institute of Technology.}
\author{C. Noel}%
 \email{crystal.noel@duke.edu}
\affiliation{%
 Department of Electrical and Computer Engineering, Duke University, Durham, North Carolina 27708, USA
 \\
 Department of Physics, Duke University, Durham, North Carolina 27708, USA
 \\
 Duke Quantum Center,
 Duke University, Durham, North Carolina 27708, USA
}%

\renewcommand{\appendixname}{APPENDIX}


\begin{abstract}
Sympathetic cooling is a technique often employed to mitigate motional heating in trapped-ion quantum computers. However, choosing system parameters such as number of coolants and cooling duty cycle for optimal gate performance requires evaluating trade-offs between motional errors and other slower errors such as qubit dephasing. The optimal parameters depend on cooling power, heating rate, and ion spacing in a particular system. In this study, we aim to analyze best practices for sympathetic cooling of long chains of trapped ions using analytical and computational methods. We use a case study to show that optimal cooling performance is achieved when coolants are placed at the center of the chain and provide a perturbative upper-bound on the cooling limit of a mode given a particular set of cooling parameters. In addition, using computational tools, we analyze the trade-off between the number of coolant ions in a chain and the center-of-mass mode heating rate. We also show that cooling as often as possible when running a circuit is optimal when the qubit coherence time is otherwise long. These results provide a roadmap for how to choose sympathetic cooling parameters to maximize circuit performance in trapped ion quantum computers using long chains of ions.
\end{abstract}

\maketitle


\section{\label{sec:intro}Introduction}



Trapped-ion quantum computers are one near-term implementation of quantum computing, where ions trapped in electric fields form chains along the electric potential's weakest axis \cite{Haeffner2008review, Bruzewicz2019}. 
A significant challenge to high-fidelity entangling gate operations in trapped ions is motional heating that causes significant gate errors and decoherence in long chains of ions \cite{MarkoAxial,Sutherland2022,Semenin2022}. 
One proposed method to combat this heating is with sympathetic cooling \cite{Itano1991a,Kielpinski2000,Mao2021,Aikyo2023}. Introducing ancillary coolant ions into the chain interspersed between the qubit ions, the Coulomb coupling between the ions can be used to cool the qubits indirectly by laser-cooling the coolants. Thus, the electronic state of the qubits is not destroyed while the chain is being cooled.


However, this strategy comes with trade-offs; firstly, adding too many coolant ions can increase the length of the chain, making it more susceptible to thermal noise due to the reduction in motional mode frequency \cite{Cetina2022, Brownnutt2015a}. In addition, gates cannot be run while ions are being cooled due to the use of ion motion to execute the gate; thus, if the chain is cooled for too long, other errors can dominate the noise budget, such as dephasing effects. 

In this paper, we present an analysis of the various parameters taken into account when cooling an ion chain using sympathetic cooling. By simulating and analyzing sympathetic cooling for trapped-ions, we find optimums for coolant placement and cooling duty cycle, and weigh the effects of system improvements on sympathetic cooling performance for long sequences of gates.

\section{\label{sec:models} Modeling chain dynamics}

We define a dimensionless potential $V(\Vec{u})$ normalized to MHz frequency units \cite{laird_thesis}, i.e. $V = U/E_0$ and $\Vec{u} = \Vec{x}/d_0$, where

\begin{equation*}
    E_0 = \frac{e^2}{4\pi\epsilon_0d_0},~~d_0 = \left(\frac{e^2}{4\pi\epsilon_0m(2\pi \times \text{1 MHz})}\right)^{1/3}
\end{equation*}



Ions are often trapped in an equally-spaced chain, allowing a single optical device with equally spaced channels such as an acousto-optic modulator or array of fibers to optically address all ions easily \cite{egan2021,Binai-Motlagh2023}. To create an equispaced chain, a quadratic potential is insufficient. We compensate by adding in a quartic term to our potential. Thus, in the direction of the chain, ions can be trapped at normalized positions $u_i$ in a symmetric quartic potential of the form:


\begin{equation} \label{eq:full_pot}
    V(\mathbf{u}) = \sum_{i} X_2u_i^2+X_4u_i^4 + \frac{1}{2} \sum_{i\neq j} \frac{1}{|u_i - u_j|}.
\end{equation}
The last term represents the Coulomb interaction between ions. Approximate equispacing of 15 ions at 4.4 $\mu$m is achieved by setting $X_2=0.00188, X_4=0.00177$. 

\begin{figure}[ht]
    \centering
    \includegraphics[width=0.5\textwidth]{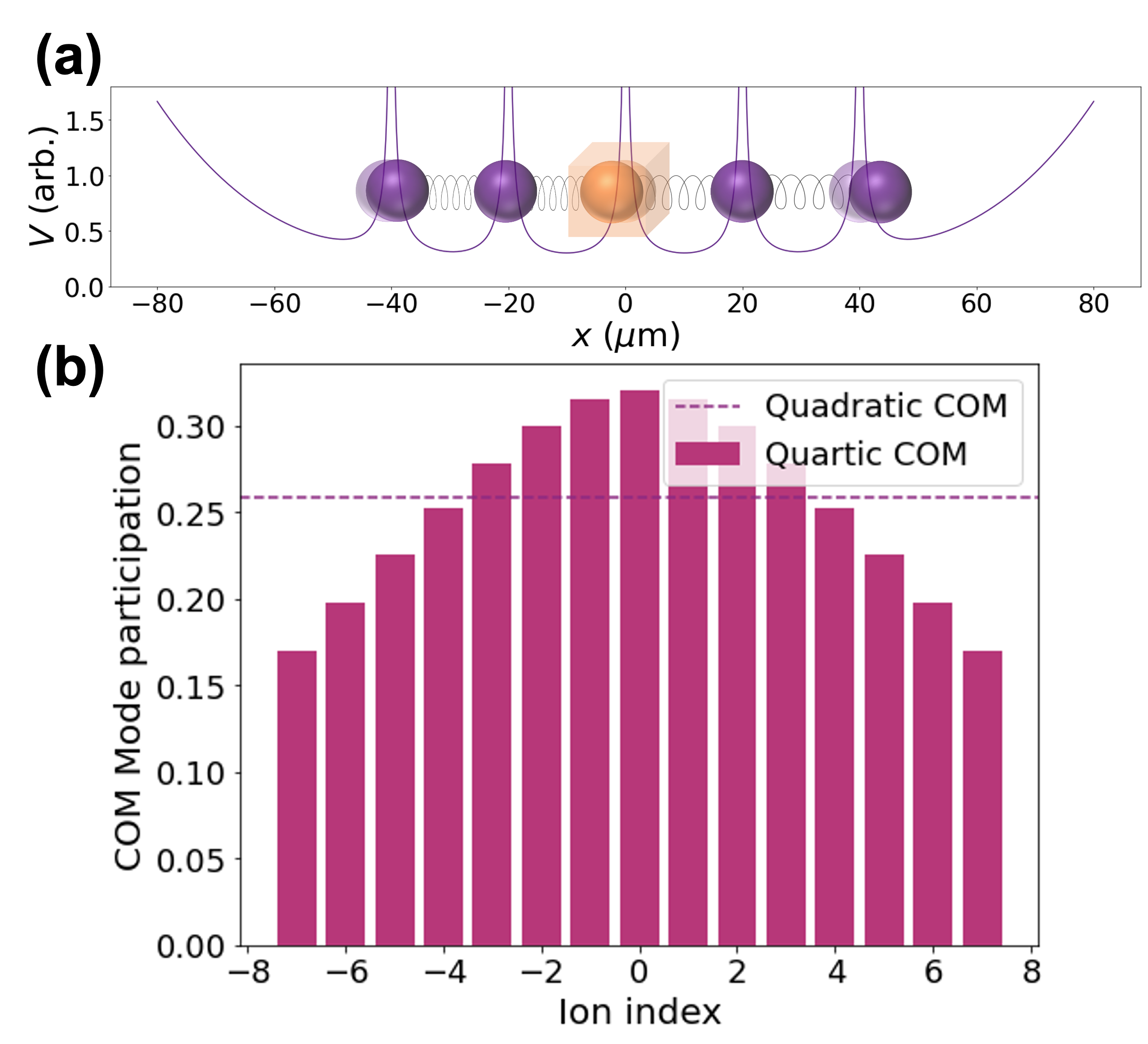}
    \caption{\textbf{Semiclassical model for trapped ions.} (a) Ions trapped in a quadratic potential can be approximated as a simple coupled oscillator system. (b) Participation factors for an equally spaced chain of 15 ions in a potential with both a quadratic and quartic term. Participation factors for a purely quadratic potential are equal across the chain (dotted line).}
    \label{fig:fig-1}
\end{figure}

\subsection{Semiclassical model}



At equilibrium, the ions are cold enough that the potential can be linearized, allowing the dynamics of the ions in the chain to be well-approximated as a set of coupled oscillators \cite{laird_thesis, Guggemos2015}.
We use Eqn.~\ref{eq:full_pot} to generate an equation of motion for this coupled oscillator system by taking the Hessian $K$ of $V(\mathbf{u})$, a set of effective spring constants and coupling coefficients between the ions. 
The solutions to such a system are well-known and form the normal modes of this chain. The lowest-frequency of these modes in the axial direction is the center-of-mass (COM) mode. In this mode, all the ions move in phase with each other.
The COM mode is generally the largest contributor to axial heating in ion chains, as low-frequency electric field noise can be approximated as a DC electric field displacing the entire chain in the same direction. 




We can model cooling by extending the semiclassical model to include a damping force \cite{Metcalf2003, Guggemos2015, Morigi2001}. The attenuation of a particular mode scales with the mode amplitude, yielding exponential decay of the mode amplitude over time. The damping matrix, $\Gamma$, represents the damping force imparted upon the ions by the cooling light.


When undamped, chains heat linearly in time, which places a bound on the number of quanta that can be cooled out of a mode. This heating is analogous to a constant forcing term in the semiclassical model. The cooling limit, $n_0$, is the ratio between a mode's heating and cooling rate.

\begin{figure}[t] 
    \centering
     \includegraphics[width=0.5\textwidth]{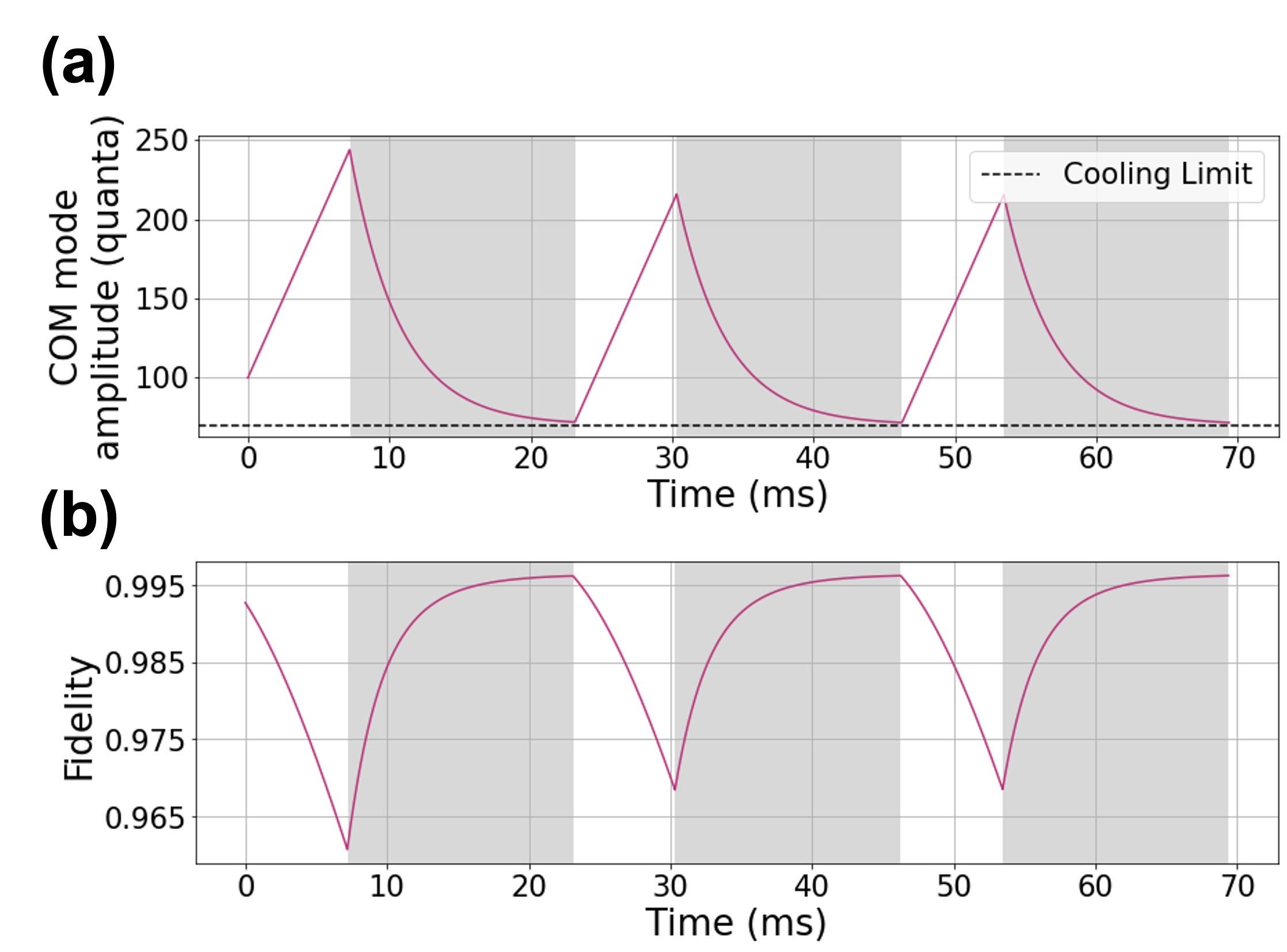}
        \caption{\textbf{Cooling dynamics.} (a) simulated COM mode amplitude (with analytical cooling limit described in section \ref{sec:models} illustrated), and (b) simulated mean XX gate fidelity with a $60\%$ duty cycle (60\% of each cycle is spent cooling the chain) over three cooling cycles. Shaded sections correspond to sympathetic cooling intervals, while unshaded sections represent intervals where gates are being run. Parameters used in this simulation include a chain length of 23 ions with $3.7~\mu$m spacing and 640~kHz cooling Rabi frequency.}
        \label{fig:dyn_fid}
\end{figure}

Thus, our final model takes the form of coupled, damped, and forced harmonic oscillators, and can be given as follows:
\begin{equation} \label{eq:dyn_eq}
    \mathbf{x}''(t) = K\mathbf{x}(t)+\Gamma\mathbf{x}'(t)
\end{equation}
By using an ansatz $\mathbf{x}(t) = \mathbf{v}e^{i\omega t}$, we can derive the eigenvalues of $K$ \{$-\omega_i^2$\} and its corresponding eigenvectors \{$\mathbf{v}_i$\}. These eigenvectors represent the modes of this oscillator system, while the eigenvalues represent their frequencies. These dynamics can be seen in Fig.~\ref{fig:dyn_fid}, with linear increase in mode amplitude when heating and exponential decay of the mode when cooling.

\subsection{Coolant placement - perturbative approach}




To find the eigenvalues of this system, we can use a perturbative approach for two reasons. Firstly, the damping force present on single ions is small relative to the mode frequencies. The timescale for COM mode cooling is on the order of milliseconds, which corresponds to a cooling rate of ~$10^2$ Hz, in comparison to the mode frequency of ~$10^5$ Hz \cite{Nop2021, laird_thesis}. Secondly, $\Gamma$ is a sparse, diagonal matrix, with only $N_C$ nonzero entries for $N_C$ coolants in the chain, each at indices $C$. As such, the eigenvalues and eigenvectors of the damped system will be close enough to those of the unperturbed system, and thus we can approximate the new eigenvectors and eigenvalues based on the original modes and frequencies of the system.

Because the cooling rate is constant for all coolant ions, we can rewrite $\Gamma = \gamma P$, where $P$ is some projector matrix with ones at specific coolant positions along the diagonal. 
Next, we use a similar ansatz to find our eigenvalues and eigenvectors. By using $\mathbf{x}(t) = \mathbf{w}e^{zt}$, we will get eigenvectors $\mathbf{w}_i$ and complex eigenvalues $z_i^2$. $\Im\{z_i\}$ corresponds to the oscillation frequency of the $i$th damped mode, while $\Re\{z_i\}$ corresponds to the cooling rate of the $i$th damped mode. 
Because we can solve this system perturbatively with respect to $\gamma$, we set $\mathbf{w}_i = \sum_{k=0}^\infty \gamma^k \mathbf{w}_i^{(k)}$, and $z_i^2 = s_i = \sum_{k=0}^\infty \gamma^k s_i^{(k)}$. In addition, $\mathbf{w}_i^{(0)} = \mathbf{v}_i$, while $s_i^{(0)} = -\omega_i^2$. We substitute the ansatz into Eqn.~\ref{eq:dyn_eq}:

\begin{align*}
    s_i\mathbf{w}_i &= \left(K + \gamma P \frac{\partial}{\partial t}\right) \mathbf{w}_i\\
\end{align*}

We can see here that the relevant perturbation is the $P\frac{\partial}{\partial t}$ term. Using this, we can find the first-order eigenvalue correction for $z_i^2 = s_i$:

\begin{align*}
    s_i^{(1)} &= \mathbf{v}_i^*(t) \left(P \frac{\partial}{\partial t}\right)\mathbf{v}_i(t)\\
    &= \mathbf{v}_i^*(t) P (i\omega_i\mathbf{v}_i(t))\\
    &= i\omega_i \sum_{k \in C} |v_{ik}|^2
\end{align*}

As such, our first-order approximation of $s_i$ gives us a similar approximation for $z_i$ and its real part $\Re\{z_i\}$:

\begin{equation*}
    z_i \approx \sqrt{-\omega_i^2 + \gamma i \omega_i \sum_{k \in C}|v_{ik}|^2}
\end{equation*}


\begin{equation}
\label{eq:cooling_rate}
    \Re\{z_i\} \approx -\frac{1}{\sqrt{2}}\sqrt{\sqrt{\omega_i^4 + \gamma^2\omega_i^2\left(\sum_{k \in C} |v_{ik}|^2\right)^2} - \omega_i^2}
\end{equation}

In the case that the modes hybridize, the mode participation vectors would be different. As such, we can perturbatively expand those in $\gamma$ as well, which we do as follows:

\begin{align}
    \mathbf{w}_i^{(1)} &= \sum_{k \neq i} \frac{1}{\omega_i^2 - \omega_k^2} \mathbf{v}_k(t) \mathbf{v}_k^*(t)\left(P\frac{\partial}{\partial t}\right)\mathbf{v}_i(t) \nonumber \\
    &= \sum_{k \neq i} \frac{i\omega_i}{\omega_i^2 - \omega_k^2} \mathbf{v}_k \mathbf{v}_k^*P\mathbf{v}_i \nonumber \\
    &= \sum_{k \neq i} \left(\frac{i\omega_i}{\omega_i^2 - \omega_k^2} \sum_{j \in C} |v_{jk}|^2 \right)\mathbf{v}_k
\end{align}

A point of note here is that the mode vectors are, by construction of $K$, orthogonal. As such, in this perturbative approximation, the only mode that the $i$th mode would hybridize to would be the $\mathbf{v}_i = \mathbf{w}_i^{(0)}$ term.

To examine the dependence of the $i$th mode's cooling rate $c_i = \Re\{z_i\}$ on its frequency, we define a variable $g_i = \frac{\gamma}{\omega_i}$. Redefining $\Re\{z_i\}$ in terms of $g_i$,

\begin{align*}
    \Re\{z_i\} &\approx -\frac{1}{\sqrt{2}}\sqrt{\sqrt{\omega_i^4 + g_i^2\omega_i^4\left(\sum_{k \in C} |v_{ik}|^2\right)^2} - \omega_i^2} \\
    &= -\frac{\omega_i}{\sqrt{2}}\sqrt{\sqrt{1 + g_i^2\left(\sum_{k \in C} |v_{ik}|^2\right)^2} - 1}
\end{align*}



To check the validity of this solution, we compare it to the computed eigenvalues of the damped system given in Eqn.~\ref{eq:dyn_eq}. The perturbative approximation was accurate to the computed eigenvalues within $\approx 10^{-6} \gamma$ for values of $\gamma$ corresponding to the range of cooling Rabi frequencies employed for Doppler cooling of trapped ions.



Furthermore, we can verify the linear approximation of the cooling rate holds for small $g_i$ by applying the binomial expansion:

\begin{align}
    \Re\{z_i\} &\approx -\frac{\omega_i}{\sqrt{2}}\sqrt{\left(1 + \frac{1}{2} g_i^2 \left(\sum_{k \in C} |v_{ik}|^2\right)^2 + O(g_i^4)\right) - 1} \nonumber\\
    &= -\frac{\gamma}{2}\left(\sum_{k \in C} |v_{ik}|^2\right) + O(\gamma^3)
\end{align}

This shows that the dependence of the cooling rate on $\gamma$ is small enough that we need only use the linear approximation of $\Re\{z_i\}$ in $\gamma$.

Numerically as well, the perturbation in $\gamma$ is generally small enough that the undamped COM mode doesn't hybridize into other damped modes significantly. In other words, the lowest-frequency damped mode has participation factors close enough to the undamped COM mode, and as such, cooling the damped COM mode will also cool the undamped COM mode. However, in the presence of higher-frequency undamped modes or larger damping parameters $\gamma$, this regime may not hold, and thus warrants further investigation.

\subsection{Cooling limit}


We can characterize the performance of a cooling scheme using the cooling limit. Because the decay rate of a mode being cooled scales with the number of quanta in the mode, the mode eventually reaches equilibrium when the heating rate $h$ is equal to the cooling rate $c$, establishing a cooling limit at a number of quanta $n_0 = \frac{h}{c}$.

To compute the cooling limit for a particular mode, we use our previous approximation for the COM mode cooling rate in conjunction with results from \cite{MarkoAxial}, which show empirically that the mode heating rate $h$ is related to axial mode frequency by a power law of the form
\begin{align}
\label{eq:heating_rate}
    h = \frac{dn}{dt} &=  D\omega_0(A_0\omega_0^{-2-\alpha} + B_0) \\
    &= (A\omega_0^{-1-\alpha} + B\omega_0)
\end{align}

In Eqn. \ref{eq:heating_rate}, $D$ is a normalization factor derived from \cite{MarkoAxial} and is dependent on chain parameters such as trapping potential and cooling Rabi frequency. Thus, in this work, we fix $A_0$ and $B_0$ as chain-independent parameters when modeling heating dynamics.In long chains, COM frequencies are generally on the order of 10$^5$ Hz \cite{MarkoAxial}. As such, so long as the axial COM frequency remains in this regime, the contribution of the $B\omega_0$ term is relatively small.

Combining these results with Eqn.~\ref{eq:cooling_rate}, we get:

\begin{align}
    \label{eq:cooling_lim}
    n_0 &= \frac{h}{c} = \frac{h}{\Re(z_0)} \nonumber \\
    &= \frac{A\omega_0^{-1-\alpha} + B\omega_0}{\frac{\omega}{\sqrt{2}}\sqrt{\sqrt{1 + g_0^2\left(\sum_{k \in C} |v_{0k}|^2\right)^2} - 1}}\\
    &\approx \frac{A\omega_0^{-2-\alpha} + B}{\frac{1}{\sqrt{2}}\sqrt{\sqrt{1 + g_0^2\left(\sum_{k \in C} |v_{0k}|^2\right)^2} - 1}}\\
    &\approx \frac{2(A\omega_0^{-2-\alpha} + B)}{g_0\left(\sum_{k\in C} |v_{0k}|^2\right)} \\
    &= \frac{2(A\omega_0^{-1-\alpha} + B\omega_0)}{\gamma\left(\sum_{k\in C} |v_{0k}|^2\right)}\end{align}

One observation from the above expression is that our cooling performance becomes much better as we increase our COM mode frequency, as this decreases our heating rate while increasing our cooling rate. In contrast, increasing the individual ion damping parameter will increase our cooling capabilities linearly, while not affecting our heating rate at all.

Secondly, we observe that the cooling rate is maximized when the coolants are placed at positions that maximize the sum of the coolants' mode participation factors, i.e. maximize the sum $\sum_{k \in P} |v_{ik}|^2$. Because the ions at the center of the chain have the highest mode participation factors in the COM mode, for \emph{any} potential, it is always optimal to place coolants at the center of the chain to cool the COM mode.

\begin{figure}
    \centering
    \includegraphics[width=0.45\textwidth]{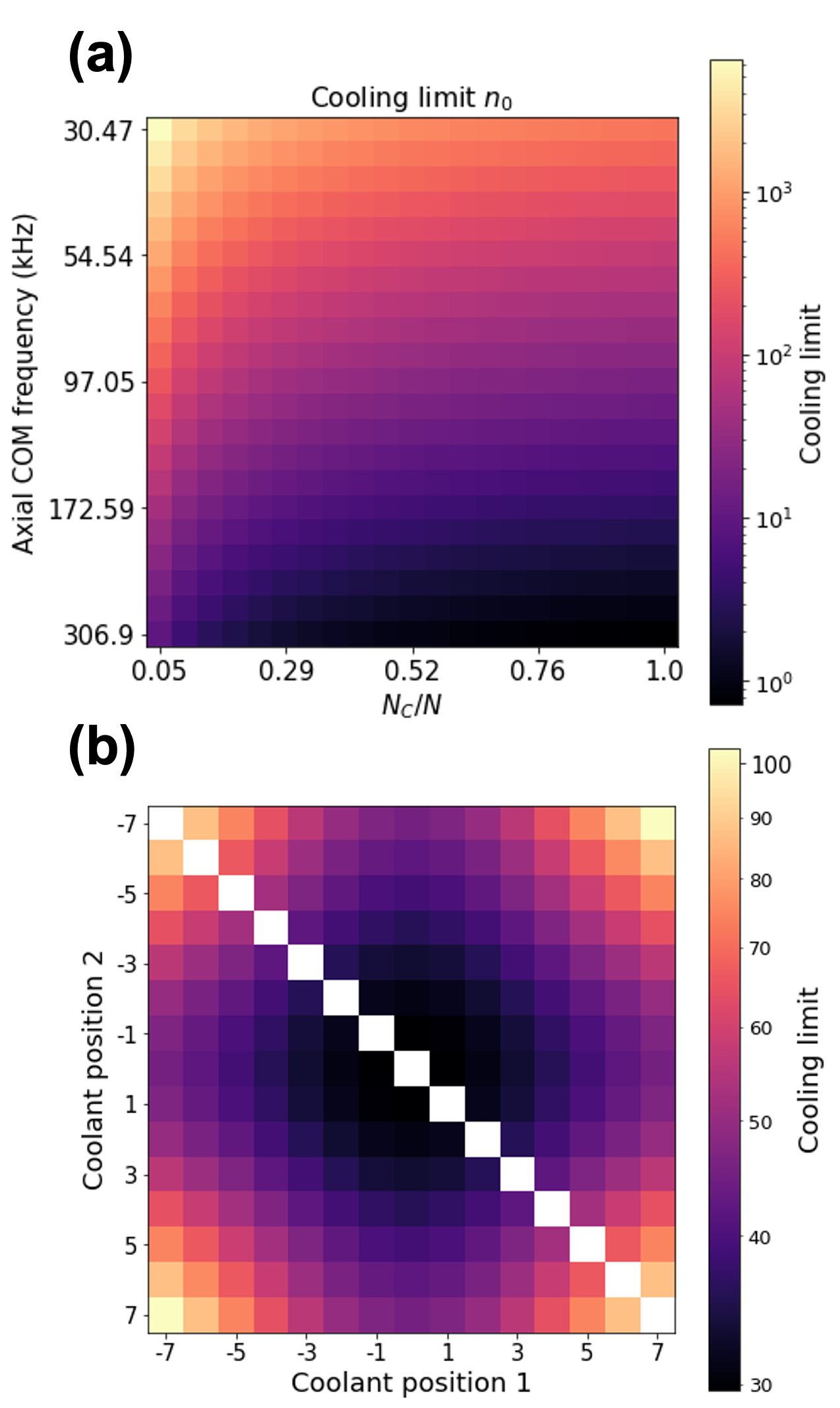}
    \caption{\textbf{Computational verification of theoretical observations.} (a) Cooling limit with respect to coolant proportion and COM frequency. The increase in cooling limit in the vertical direction corresponds with an increase in heating rate due to a lower COM frequency, while the decrease in cooling limit in the rightward direction corresponds to a larger number of coolants in the chain. The cooling limit plateaus as the chain fills with coolants. (b) Simulated COM mode cooling rates for all possible configurations of two coolants in a 15-ion chain. Optimal placement is found with ions placed at positions (0, 1) or (-1, 0), in good agreement with our theoretical observation that optimal cooling performance occurs with ions placed at the center of the chain.}
    \label{fig:perf-and-lim}
\end{figure}

Although we don't have an analytical expression for $\omega_0$ or $v_{0k}$ in a quartic potential, the highest mode participation factors for the COM mode in a quartic potential come from the ions at the center of the chain (see Fig.~\ref{fig:fig-1}), and thus we know that the optimal ion positions must be at the center of the chain.

We also use computational methods to verify the accuracy of our analytical observations by examining the correspondence between the perturbative definitions of the cooling and the computationally simulated cooling limit. Fig.~\ref{fig:dyn_fid}(a) shows the accuracy of the cooling limit. Our analytical results show good agreement to the simulated dynamics, accurately estimating the limit of a particular cooling scheme given its geometry. In addition, we also found that the perturbative approximation of the cooling rate only had a relative error of $< 2.5 \times 10^{-4}$ for all $\gamma < 10^{-3}$.

We computationally verify our observation that optimal coolant placement is at the center of a chain by simulating every possible configuration of two coolant ions in a 15-ion chain, seen in Fig.~\ref{fig:perf-and-lim}(b). Cooling performance was maximized with centrally-placed coolants (i.e. at positions (-1, 0) or (0, 1)), yielding a cooling limit of 29 quanta. This is in good agreement with our theoretical observations, as configurations that perform best (i.e. minimize the cooling limit) are centrally placed, while configurations that perform worse place coolant ions towards the edge of the chain.

However, this perturbative approximation will likely not hold at higher values of $\gamma$ due to hybridization of the undamped COM mode into other higher-order modes of damped motion that arise in long ion chains. In the small $\gamma$ regime, the perturbation does not significantly change the modes of the damped system when compared to those of the undamped system. As such, each undamped mode largely couples to its damped counterpart, including the undamped COM mode.

Finally, we observe that many of the quantities in the above expression can be determined analytically for a quadratic potential. For a normalized quadratic potential given by $V(u_i) = X_2 u_i^2$, our normalization gives $\omega_0 = (2\pi \times 1~\text{MHz}) \sqrt{X_2}$, while $v_{0k} = N^{-1/2}$ for all positions in an $N$-ion chain \cite{Morigi2001}. The latter reduces the sum $\sum_{k \in P} |v_{ik}|^2$ to $\frac{N_C}{N}$. This provides an analytical cooling limit for any configuration of an ion chain trapped in a quadratic potential.

However, this also provides us with an analytical upper-bound for a quartic potential. In the limit where a few coolants are placed optimally at the center of a chain trapped in a quartic potential, the higher mode participation factors at the center of the chain ensure that the sum $\sum_{k \in P} |v_{ik}|^2 > \frac{N_C}{N}$, while the inclusion of a nonzero $X_4$ term will only make the potential tighter, thus causing a higher COM frequency. As such, because we can show that the analytical quadratic heating rate is higher than its quartic counterpart and the analytical quadratic cooling rate is lower than the quartic cooling rate, the quadratic cooling rate can act as a soft upper bound on the cooling limit of a corresponding quartic chain.

\section{\label{sec:sim} Simulation and Analysis} 

Now we apply the above theoretical model to find optimal cooling parameters for long-chain trapped-ion quantum computers through computational simulation by maximizing mean gate fidelity over time.
We first compute the eigenfrequencies and eigenmodes of the system given in Eqn.~\ref{eq:dyn_eq} in both the undamped ($\Gamma = 0$) and damped ($\Gamma = \gamma P$) regimes for a particular cooling configuration. 
We then use the eigenvalues of Eqn.~\ref{eq:dyn_eq} to compute the cooling rate of the COM mode, while using Eqn.~\ref{eq:heating_rate} to compute the COM heating rate. These rates are subsequently used to simulate COM mode dynamics, as shown in Fig.~\ref{fig:dyn_fid}(a).
Finally, we use results found in \cite{Cetina2022} to compute XX gate fidelity from the resultant mode dynamics, seen in Fig.~\ref{fig:dyn_fid}(b).

Cooling tradeoffs are evaluated by analyzing the number of gates run per cooling cycle and the amount of time used for cooling per cycle. Using these ideas, we define a ``cooling duty cycle", characterized by the fraction of the cycle used for cooling.

To evaluate cooling performance, we examine mean XX gate fidelity and total circuit fidelity. Mean XX gate fidelity examines the average gate fidelity over the circuit, while total circuit fidelity establishes a worst-case performance bounds. In particular, for a circuit with $M$ gates, where the $k$th gate is run on the interval $[t^{(k)}_{start}, t^{(k)}_{stop}]$, mean XX gate fidelity $\langle F\rangle$ and total gate fidelity $F_{total}$ are defined as follows 
\footnote{It can be shown analytically that these metrics are equivalent under ranking, as mean XX gate fidelity corresponds to an arithmetic mean, while total gate fidelity corresponds to a geometric mean. Thus, for a fixed number of XX gates, the two metrics are equivalent, as the total fidelity is the exponentiation of the mean fidelity in logscale, i.e. 

$$F_{total} = \exp\left(\sum_{k=1}^M \ln F(t_{start}^{(k)})\right)$$

However, there may be other preferred metrics, such as to simulate an entire circuit when possible.}:

\begin{align}
    \langle F\rangle &= \frac{1}{M} \sum_{k=1}^M F(t^{(k)}_\text{start})\\
    F_{total} &= \prod_{k=1}^M F(t^{(k)}_\text{start})
\end{align}




Gate fidelity depends on the shape of the individual addressing beams used in the specific system, in addition to the COM mode frequency. For these properties, we utilize parameters used \emph{in situ} in \cite{MarkoAxial}. We examine two separate sets of parameters to optimize: \emph{chain parameters} such as the number of coolants, and \emph{cooling parameters} such as cooling duty cycle. 

As a proof of concept, the simulation was used to find optimal coolant ion configurations and cooling schedules for a quantum algorithm given in \cite{KeeAlgo} run on the trapped-ion quantum computer specified in \cite{MarkoAxial}. 

\subsection{Parameters}

The parameters for the initial simulation were largely derived from \cite{MarkoAxial}. Heating rates for the ion chain were derived using the relation in Eqn.~\ref{eq:heating_rate}
with $\alpha = 0.8$, $A_0 \approx 8.2\times 10^{17} \text{ s}^{-1} \times (1 \text{ rad/s})^{2+\alpha}$, and $B_0 = 0.9\text{ s}^{-1}$, as these parameters yielded good agreement with experimental data. We conduct simulations using cooling Rabi frequencies of 180, 275, and 640~kHz. The corresponding cooling rates were also determined to be $\gamma$ = $1.387 \times 10^{-5}$, $3.468 \times 10^{-5}$, and $5.328 \times 10^{-5}$ respectively, as these parameters also gave good agreement with the cooling dynamics described in \cite{MarkoAxial}.

Because single-qubit gates are faster than two-qubit gates and their fidelities are not as dependent on axial heating, we exclude them from analysis. For this case study, we optimize 500 two-qubit entangling gates. Each gate has a duration of 250 $\mu$s, yielding a minimum circuit time of 125 ms without any cooling. The gates are run sequentially on 14 qubits. Each simulated ion chain has $N = N_C + N_Q + 2$ ions, where the number of $^{171}$Yb$^+$qubits, $N_Q$, remains fixed at 14, while the number of $^{172}$Yb$^+$ coolants, $N_C$, is varied and the performance of each configuration is analyzed. In this optimization, 2 extra ``endcap" ions are assumed to maintain equal spacing of the inner ions. 

\subsection{Chain optimization}

\begin{figure*}[ht]
    \centering
    \includegraphics[width=\textwidth]{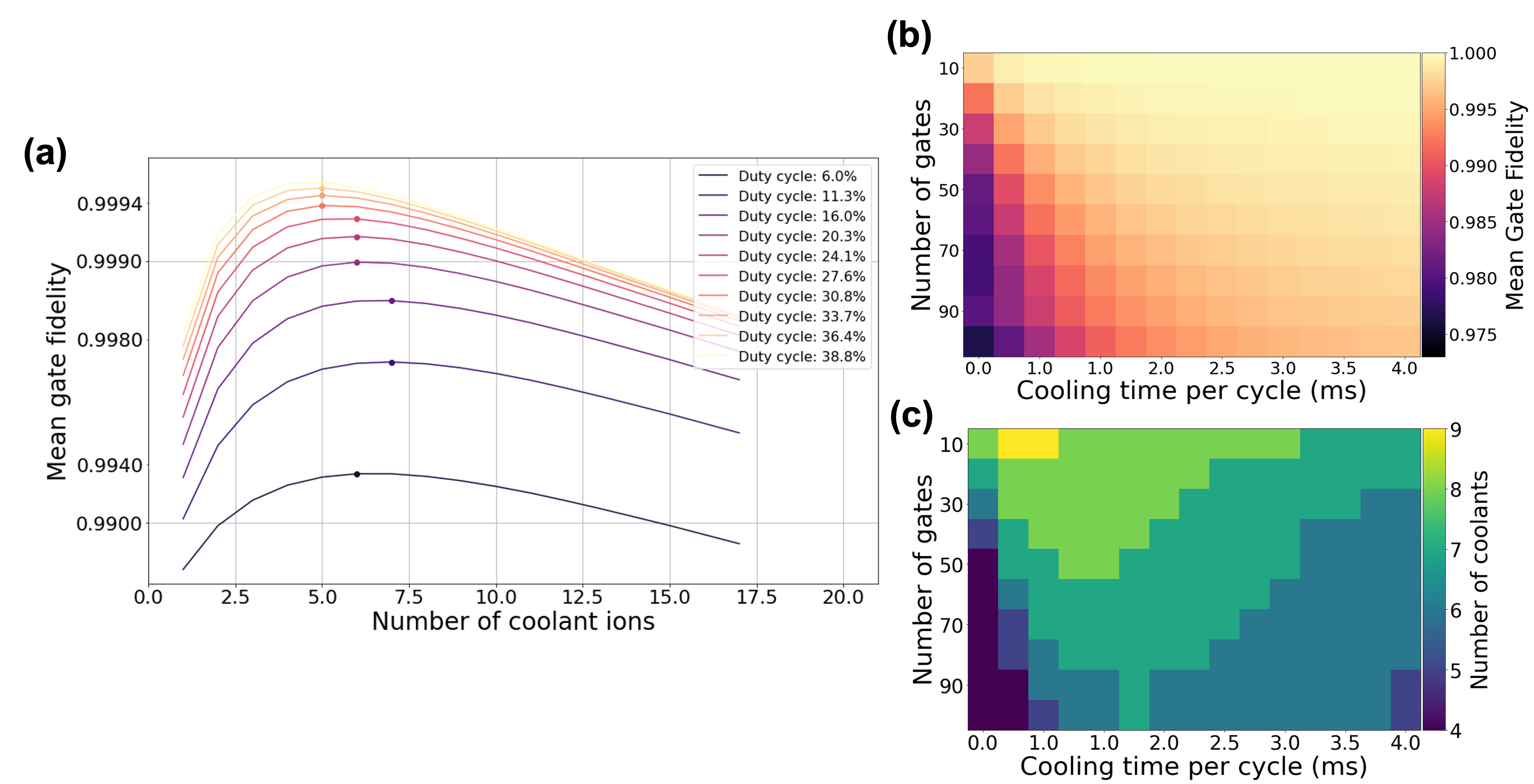}
    \caption{\textbf{Chain parameter dynamics.} (a) Mean gate fidelity for an XX gate for different duty cycle parameters as the number of coolants in the chain is increased. The trade-off between cooling and heating rates is shown here, with fewer coolants yielding too low of a cooling rate and too many coolants increasing heating rates. (b) Optimal gate fidelity w.r.t. gates per cycle, cooling duty cycle, and number of coolants in the chain. (c) The corresponding number of coolants required to achieve the maximal fidelities found in (b).}
    \label{fig:chain-opt}
\end{figure*}

To optimize the chain parameters, we run the same cooling simulation while varying the number of coolants in the chain, and thus the frequency of the axial center-of-mass mode. All coolant ions in the simulation are placed in the center, as we showed in section \ref{sec:models} that ion placement in the center of the chain is optimal for all cooling configurations.


The impact of COM mode frequency in a fixed potential (and, by proxy, chain length) has been examined at length in \cite{MarkoAxial}. Thus, in this study, we examine cooling performance through the lens of mean XX gate fidelity and how it is impacted by the number of coolants and chain length. Because these parameters are determined by the physical characteristics of the trapped-ion system, these results are largely dependent on the number of ions used to run a circuit, and are independent of circuit length and gate type.

We first compute the mean gate fidelity of a circuit run on the chain as we increase the number of coolants in the chain with a fixed number of qubits. As seen in Fig.~\ref{fig:chain-opt}(a), the optimal number of coolants depends on the cooling duty cycle.
Having fewer than the optimal number of coolants in the chain causes the coolants to not couple strongly enough to the COM mode, leading to insufficient cooling of the mode, lowering gate fidelity. However, more than 6 coolants causes the COM mode frequency to decrease due to chain length (as shown in \cite{MarkoAxial}), and thus increases the overall heating rate of the COM mode, making the chain harder to cool. 
As the cooling time for each cooling cycle is increased, the optimal number of coolants varies. Thus, the optimal chain configuration is largely dependent on cooling parameters, and may be determined computationally for fixed cooling cycle parameters.
For the purposes of this study, it's sufficient to assume a fixed electric potential for ion trapping. However, below, we also briefly examine how changing aspects of the trap potential (in particular, the COM mode frequency) impacts cooling performance. 

Rather than adding coolants to the chain for a fixed number of qubits, we examine the ratio of coolants to total ions in the chain. In so doing, we examine the effect of increasing the cooling rate of the mode by maximizing the total mode participation over all coolants (i.e., maximizing $\sum_{k \in P} |v_{0k}|^2$ in Eqn. \ref{eq:cooling_rate}).
We test 21-ion chains with differing COM frequencies as we add coolants starting at the center of the chain, subsequently assessing their cooling performance. Fig. \ref{fig:perf-and-lim}(a) shows the corresponding cooling limit of a chain for a fixed COM frequency and coolant filling fraction. The theoretical prediction for our cooling limit holds well here, with the cooling limit decreasing as the COM frequency increases. In addition, the cooling performance plateaus as the coolant filling fraction approaches 1. This is indicative of the trends observed in Fig. \ref{fig:chain-opt}, in which cooling performance plateaus and eventually decreases as chain length increases.


\subsection{Cooling optimization}

\begin{figure*}[t]
    \centering
    \includegraphics[width=0.9\textwidth]{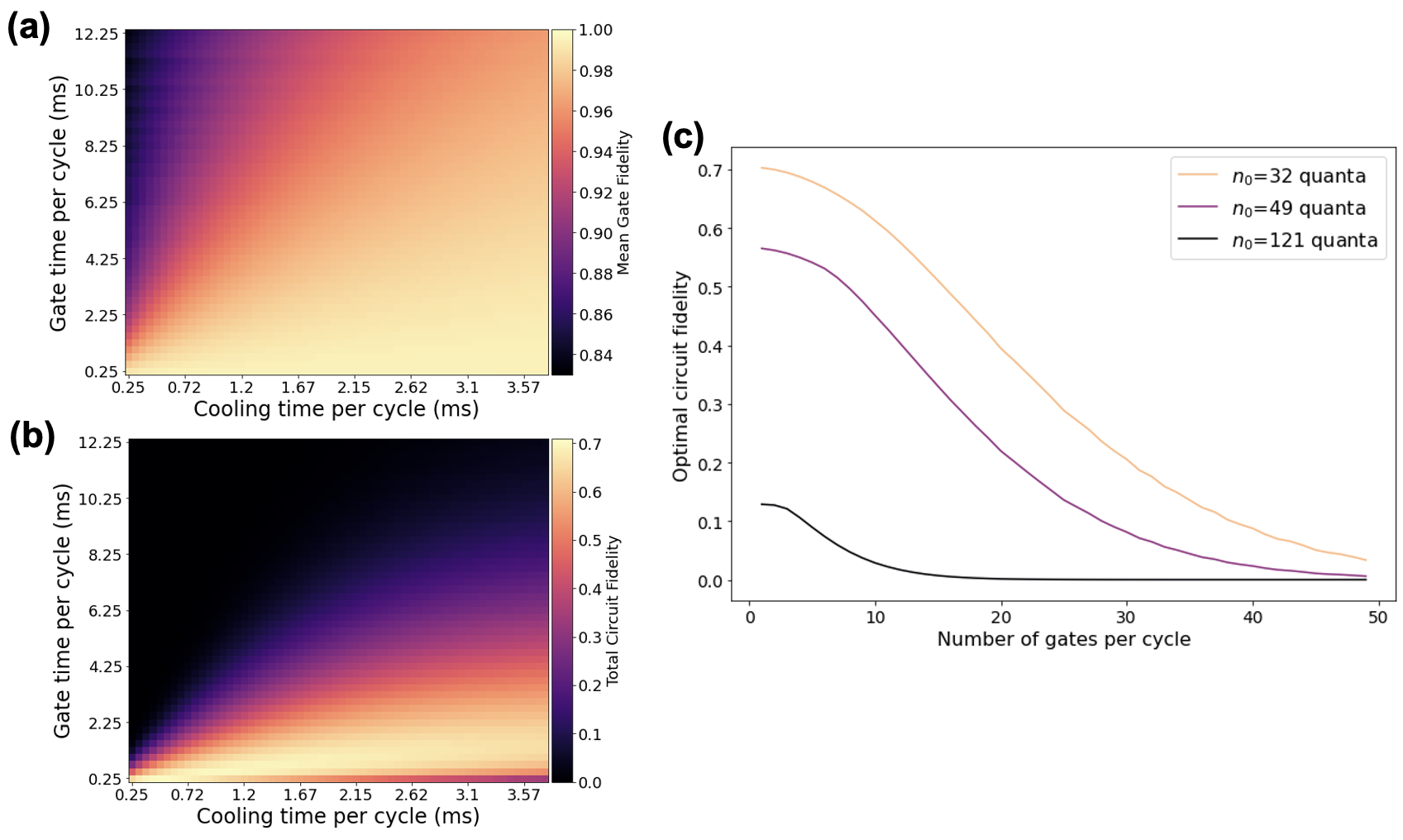}
    \caption{\textbf{Cooling parameter dynamics.} (a) Mean gate fidelity for various choices of gate time per cycle and cooling time per cycle. (b) Total circuit fidelity for various choices of gate time per cycle and cooling time per cycle. (c) Optimal circuit fidelity for each examined Rabi frequency ($\Omega_C$=640 kHz, 275 kHz, 180 kHz) as the number of gates per cooling cycle is increased. The optimal configuration for all examined Rabi frequencies occurs when only one gate is run per cycle.}
    \label{fig:cooling-opt}
\end{figure*}


Next, we optimize the cooling time and duty cycle. Increasing total duration of the circuit by increasing cooling time can mitigate motional errors, but can incur errors dominant at longer timescales, such as dephasing and depolarizing noise. On the other hand, cooling infrequently and for short times allows for circuits to run quickly (avoiding dephasing errors), but runs into motional noise, which is dominant at shorter timescales. Thus, we present an analysis of cooling cycle frequency and duration as it pertains to circuit performance.
For cooling optimization, we examine gate fidelity while varying cooling time per cooling cycle and number of gates per cooling cycle. 
Incorporating a dephasing T2 time of 0.5~s into the simulation, we aim to find a balance in the cooling duty cycle that optimizes circuit fidelity. 

For a cooling Rabi frequency of 640 kHz, we find that an optimal cooling duty cycle consists of one gate and a $\approx$75\% cooling duty cycle. This results in a total circuit fidelity of 0.73 and a mean gate fidelity of 0.9993. As seen in Fig.~\ref{fig:cooling-opt}, as cooling duration is increased, average gate fidelity increases quickly when the duty cycle is short enough to address motional errors before slowly dropping off due to T2 times. In addition, the fidelity decreases as the number of gates per cycle is increased, as increasing the number of gates per cycle leads to more time for motional error to accrue.
In all cases, we observed that optimal fidelity was achieved when cooling after every gate run in the circuit, as seen in Fig.~\ref{fig:cooling-opt}. This corresponds to the fact that $F \propto \frac{1}{\sqrt{1 + n^2}}$ \cite{MarkoAxial}. Thus, for small $n$ ($n \approx 0$), $dF/dn \approx 0$, meaning that by running only one gate at a time, the errors due to heating do not have time to manifest.




However, the exact cooling parameters found through optimization vary. At a Rabi frequency of 180 kHz, optimal fidelity is achieved at a cooling duty cycle of 88.14\% (1673 $\mu$s cooling time per gate), while at 275 kHz, we get a duty cycle of 76.31\% (724 $\mu$s cooling time per gate), and at 640 kHz, we get a duty cycle of 68.41\% (487 $\mu$s cooling time per gate). Thus, although our optimal mean-gate fidelity doesn't change significantly, our system becomes much more efficient with hardware improvements.

It is important to also note that in this analysis we do not account for photon recoil heating of the radial modes. It was shown in \cite{MarkoAxial} that these modes could also be cooled using the same methods as the COM mode, and in a similar amount of time. Therefore, we can assume that cooling the radial modes would add overhead to the cooling process, although the exact optimal scheme for cooling all modes is outside the scope of this paper. During this added cooling overhead, we neither run gates nor cool. We assume that an optimal cooling scheme could be constructed empirically to ensure all modes are close to the cooling limit at the end of the cooling time, such as alternating which mode is cooled over time.

When accounting for photon recoil, optimal fidelity was still achieved when cooling as often as possible. The amount of time required for optimal performance changed slightly: a Rabi frequency of 180 kHz yielded optimal results at a duty cycle of 84.82\% (1258 $\mu$s of cooling time per cycle), a 275 kHz Rabi frequency yielded an optimal duty cycle of 70.84\% (547 $\mu$s of cooling time per cycle), while at 640 kHz, a duty cycle of 65.54\% was found to be optimal (428 $\mu$s of cooling time per cycle).
Each of these optimal duty cycles was found to be slightly lower than the corresponding results that do not account for radial cooling time. This descrease is most likely due to the fact that extra radial cooling leads to longer circuits, which causes dephasing errors to become more prevalent in the circuit error budget. Reducing the cooling time slightly can trade off heating performance for a shorter circuit, and thus smaller dephasing errors. This leads to generally increased performance for configurations with smaller cooling duty cycles.








\section{Outlook}
As discussed in \cite{Cetina2022}, the operation of long ion chains for quantum computation is primarily limited by motional heating along the axis of the chain. A solution is proposed to used sympathetic coolants, and in this study we expand that idea and optimize coolant placement and number to maximize circuit fidelity given certain experimental parameters. 
We have formulated a first-order approximation of cooling dynamics to analytically show that coolant placement at the center of the chain is optimal, placing an upper bound on the cooling limit of the COM mode of an ion chain.
Our theoretical findings show that improvements in heating rate allow for a much more significant decrease in cooling limit, and therefore optimal circuit fidelity.
Unsurprisingly, we find that the number of coolants needed for a chain must strike a balance between increasing axial heating due to a decrease in COM mode frequency and increasing cooling rate due to stronger coupling to the COM mode.


We have also examined these results through computational simulation, showing that these theoretical findings are consistent with numerical models. Our numerical studies have allowed us to formulate a set of best practices for cooling optimization, showing that cooling as often as possible is optimal for any choice of cooling duty cycle. 
In addition, we demonstrate with a case study how to numerically find the best parameters for ideal performance of a particular quantum circuit. 
We aim to use these results to computationally optimize coolant configuration and cooling duty cycle depending on circuit specifications.

This study examines the influence of system improvements on the performance of various cooling schemes; in particular, this study examines how improvements in cooling power and cooling rate can lead to improvements in circuit fidelity. However, further study is required regarding improvements of other aspects of the device, such as decreasing heating rates \cite{Hite2012,Daniilidis2014} or using open-loop control methods such as dynamical decoupling to achieve a longer T2 time \cite{Viola1998,Biercuk2009}.\\

\section{Acknowledgments}

The authors are grateful to Marko Cetina for his review of the manuscript, and to Chris Monroe, Or Katz and Debopriyo Biswas for fruitful conversations. C.N. acknowledges support from the NSF under QLCI: Center for Robust Quantum Simulation OMA-2120757. A.P. acknowledges support from the A.B. Duke Scholarship and the Campbell L. Searle Graduate Fellowship.

\bibliography{apssamp,Noel_mendeley_abbrv}



\end{document}